\documentclass[10pt,twocolumn,preprintnumbers,superscriptaddress,nofootinbib,aps,prd]{revtex4-2}

\usepackage{graphicx}
\usepackage{amsmath}
\usepackage{srcltx}
\usepackage[utf8]{inputenc}
\usepackage[colorlinks]{hyperref}
\usepackage{url}
\usepackage[pdftex]{xcolor}
\usepackage{times}
\usepackage{amssymb}
\newcommand{\vev}[1]{\langle {#1} \rangle}

\newcommand{\gsim}{\gtrsim}

\newcommand{\ord}[1]{\mathcal{O}{(#1)}}
\newcommand{\beq}{\begin{equation}}
	\newcommand{\eeq}{\end{equation}}
\newcommand{\bea}{\begin{eqnarray}}
	\newcommand{\eea}{\end{eqnarray}}

\newcommand{\invfb}{fb$^{-1}$}

\newcommand{\appropto}{\mathrel{\vcenter{
			\offinterlineskip\halign{\hfil$##$\cr
				\propto\cr\noalign{\kern2pt}\sim\cr\noalign{\kern-2pt}}}}}

\definecolor{orange}{rgb}{0.9,0.4,0.0}

\begin{document}
	
\pagestyle{plain}

\title{Long-Lived Dark Hadrons at the Electron-Ion Collider}	

\author{Hooman Davoudiasl}
\affiliation{
High Energy Theory Group, Physics Department, \\
Brookhaven National Laboratory, Upton, NY 11973, USA
}
\author{Hongkai Liu}
\affiliation{
High Energy Theory Group, Physics Department, \\
Brookhaven National Laboratory, Upton, NY 11973, USA
}             
\author{Ethan T. Neil}
\affiliation{Department of Physics, University of Colorado, Boulder, Colorado 80309, USA}    
	
	
\begin{abstract}
We study a dark non-Abelian gauge sector with GeV-scale confinement. The dark sector is assumed to couple only feebly to the Standard Model, while its low-energy spectrum may contain long-lived flavor-diagonal dark pions. Signals of these states are particularly well suited to the Electron-Ion Collider (EIC), where the absence of a hard trigger requirement and the capability to record soft final-state particles offer a complementary probe of dark hadronization dynamics. We present a benchmark portal construction, discuss the mixing between an axion-like mediator and dark pions, and identify the resulting displaced-decay signature.

\end{abstract}
\maketitle

\section{Introduction}
Dark sectors with their own confining gauge dynamics are a broad and motivated class of extensions of the Standard Model (SM).  Although they can appear in other contexts, confining dark sectors can be especially interesting as models of composite dark matter \cite{Cline:2021itd,Asadi:2026mip}.  Composite dark matter for which the dark matter abundance is related to an asymmetry in dark baryon number, in close analogy with ordinary baryonic matter, can give a natural explanation for the similar abundance of baryonic and dark matter observed in the universe.   In such a circumstance, one would expect the mass of dark matter to be close to the nucleon mass of $\sim 1$~GeV, which may be generated by dark confining dynamics \cite{Davoudiasl:2012uw,Petraki:2013wwa,Zurek:2013wia}.  If the confinement scale is near the GeV scale, the dark sector can contain a spectrum of dark mesons and baryons whose masses are low enough to be produced at intensity-frontier experiments, while their couplings to the SM remain small enough to evade existing prompt searches.  

The EIC offers a particularly useful environment for probing this class of models, through the production of ``dark showers", experimental signatures that exhibit effects from hadronization in the confining dark sector (see Refs.~\cite{Han:2007ae,Baumgart:2009tn,Falkowski:2010cm,Cohen:2015toa,Schwaller:2015gea,Cohen:2017pzm,Knapen:2021eip,Albouy:2022cin,Cohen:2023mya,Cheng:2024hvq,Cheng:2024aco}; for other explorations of dark showers at BaBar and Belle-II specifically, see \cite{Bernreuther:2022jlj,Lu:2025cty}.)   In contrast with hadron collider searches that often rely on energetic objects for triggering, the EIC data can have comparatively soft particles~\cite{Adkins:2022jfp}.   This opens sensitivity to dark showers containing multiple low-energy final states. The signature emphasized here is the production of dark quarks through a light portal state, followed by dark showering and hadronization into dark pions. A subset of neutral dark pions may decay displaced from the interaction point, leading to events with multiple displaced vertices.

We consider a hidden confining gauge sector with gauge group
SU(3)$_D$ and a confinement scale in the GeV range. The field content may include light dark quarks charged under the dark color group but neutral under SM color. Such a sector appears naturally in several frameworks, in particular those where the visible (SM) and dark sectors are related by an approximate $\mathbb{Z}_2$ parity; see for example Refs.~\cite{Chacko:2005pe,Chacko:2005un,Falkowski:2006qq,Chang:2006ra,Craig:2014aea,Barbieri:2015lqa,Chacko:2016hvu,Chacko:2018vss,Batell:2022tif}.

The low-energy spectrum depends strongly on the number of light dark quark flavors. If there are no light dark quarks, the low-energy states are dark glueballs, which can mix with SM scalar states through a Higgs portal~\cite{Juknevich:2009gg,Batz:2023zef}. With one light flavor, the spectrum does not contain the same pattern of pseudo-Nambu-Goldstone bosons as ordinary two-flavor QCD. With two or more light flavors, approximate chiral symmetries give rise to dark pions. For $N_f$ light flavors, the number of light pseudoscalar mesons is generically $N_f^2-1$, up to explicit symmetry breaking and anomaly effects. This work focuses on the case of $N_f = 2$, and neutral dark pions are among the lightest dark hadrons.\footnote{Here and elsewhere in this work, by ``neutral" and ``charged" dark pions we mean the dark iso-spin counterparts to the SM pions; no dark electric charge is implied.} If their decay width is small, they can travel macroscopic distances before decaying. The displacement is controlled by the proper lifetime and by the boost inherited from the dark shower. The event-level probability to observe a signal therefore depends on both the production kinematics and the dark-hadron spectrum.  

The benchmark model we consider here is not intended on its own as a unique ultraviolet completion of new physics, but rather as a minimal realization of the physics of long-lived hadron decays at the EIC.  In contrast to other searches at energy-frontier hadron colliders like the LHC, the EIC is well-positioned to probe the details of dark confinement near the GeV scale.  For other searches for long-lived new particles at the EIC, see \cite{Batell:2022ogj,Davoudiasl:2023pkq,Balkin:2023gya,Davoudiasl:2025rpn,Balkin:2025rtc,Balkin:2026whv}.

\section{Basic Model}
The dark sector is connected to the SM  through a complex scalar $\Phi$ whose vacuum expectation value $\vev{\Phi}=v_\Phi/\sqrt{2}$ sets the dark-sector
mass scale,
\begin{equation}
  \Phi =
  \frac{1}{\sqrt{2}}\,(v_\Phi + S)\,
  \exp\!\left(i\frac{a}{v_\Phi}\right).
\end{equation}
In the above, $S$ is the radial scalar and $a$ is the Goldstone mode corresponding to the spontaneous breaking of a global $U(1)$ symmetry via $\vev{\Phi}\neq 0$.  
The leading interactions with the SM electron and the dark quarks
$\chi_i$ with $i=1,2$ can be written as 
\begin{equation}
  \mathcal{L}_{\rm portal}
  =
  \frac{1}{M}\,\Phi H \bar{L} e_R
  +
  y_{\chi_i} \Phi \bar{\chi}_{i,L}\chi_{i,R}
  + {\rm h.c.},
\end{equation}
where $H$ is the SM Higgs doublet, $L$ is the lepton doublet,
$e_R$ is the right-handed electron, and $M$ denotes the heavy scale that
generates the effective electron portal. After electroweak and dark-sector
symmetry breaking, with $\langle H\rangle = v/\sqrt{2}$ and $\langle \Phi\rangle = v_\Phi/\sqrt{2}$, the contribution to fermion masses
are
\begin{equation}
  \Delta m_e = \frac{v v_\Phi}{2M},
  \qquad
  m_{\chi_i} = \frac{1}{\sqrt{2}}\,y_{\chi_i} v_\Phi.
\end{equation}
In the above, $\Delta m_e$ is a correction to the electron mass $m_e$ and $m_{\chi_i}$ is the the mass of $\chi_i$.\footnote{We have implicitly assumed that there is another contribution to $m_e$ from the dimension-4 Higgs coupling and achieving the measured electron mass may entail some mild ($\gsim 20\%$) tuning for parts of our parameter space.}  At low energies, the portal interaction can be written as 
\begin{equation}
  \mathcal{L}_{\rm int}
  \supset
  - g_{S\chi_i}S \bar{\chi}_i \chi_i
  - g_{Se} S \bar{e} e
  - i g_{a\chi_i} a \bar{\chi}_i \gamma_5 \chi_i
  - i g_{ae} a \bar{e}\gamma_5 e ,
\end{equation}
where the effective couplings to electrons and dark quarks are
\beq
g_{Se} = g_{ae} = \frac{\Delta m_e}{v_\Phi},\quad g_{S\chi_i} = g_{a\chi_i} = \frac{m_{\chi_i}}{v_\Phi}
\eeq
The masses of $S$, $a$, and the dark hadronic scale are treated as
free phenomenological parameters. This allows us to separate production, decay, and hadronization effects in the simulation.

\section{Analysis}  

After symmetry breaking in the dark sector, the scalar and pseudoscalar mediators can be produced at the EIC and decay back into dark quark pairs,
\begin{equation}
  e A \to e A S/a, \qquad S/a \to \bar\chi \chi ,
\end{equation}
where $A$ denotes the ion with atomic number $Z$. The coherent production cross section is enhanced by a factor of $Z^2$. The dark quarks then shower and hadronize into
dark mesons. In this work, we focus on the production of dark quarks via the scalar field $S$.   Since $S$ is heavier than $a$, it sets the available energy for the dark shower, while the dark pion mass controls the multiplicity and boost distribution of the long-lived neutral states. For a two-flavor dark sector with approximate SU(2) flavor symmetry, the dark pion multiplet contains one neutral and two charged states.  We fix the mass ratio in the mass spectrum
\beq
m_{\rho_D} = 2.0 \Lambda_D,\ f_D = 0.34\Lambda_D,\ m_{\rho_D} > 2 m_{\pi_D},
\label{eq:mass}
\eeq
such that the dark vector $\rho_D$ mesons can decay into two dark pions promptly via $\rho_D^0\to\pi_D^0\pi_D^0, \pi_D^+\pi_D^-$, $\rho_D^\pm\to\pi_D^0\pi_D^\pm$.  In SU(3) gauge theory with pion masses from the physical point up to $m_{\rho_D} / 2$, the value of the ratio $m_{\rho_D} / f_D \approx 5.75$ and is roughly constant \cite{DeGrand:2019vbx}, corresponding to our choice of $m_{\rho_D} / f_D$. The heaviest dark quark mass is fixed to be $m^2_{\pi_D}/m_{\rho_D}$.
\begin{figure}[t]
	\centering
\includegraphics[width=\columnwidth]{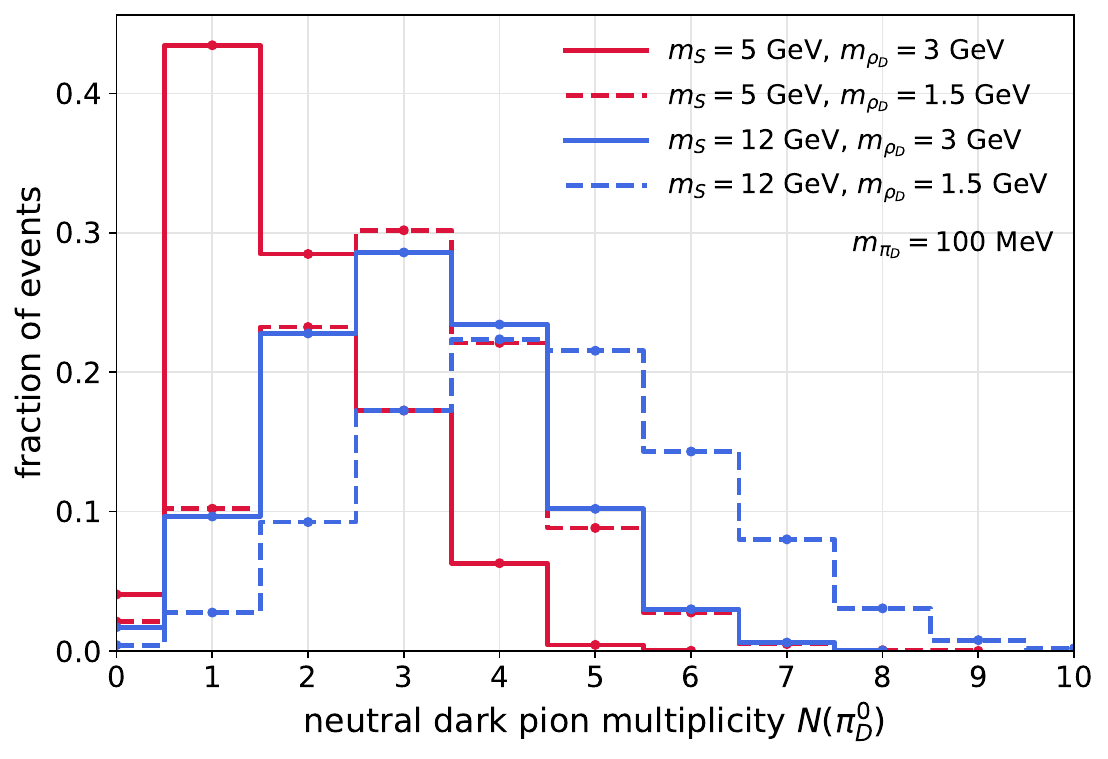}
\caption{Neutral dark pion multiplicity distributions for the four benchmark choices of $(m_S,m_{\rho_D})$, with $m_{\pi_D}=100~\mathrm{MeV}$.}
\label{fig:mul}
\end{figure}

In the following analysis, we consider four representative benchmark scenarios, corresponding to $m_S = 5,\ 12~\mathrm{GeV}$ and $m_{\rho_D} = 1.5,\ 3~\mathrm{GeV}$. 
These choices span both lighter and heavier scalar mediators, as well as two representative dark-vector mass scales.\footnote{We note that a simple scaling based on SM hadron mass ratios suggests that the lightest dark nucleon in our example would have a mass $\sim 1.8$~GeV, which avoids neutron star bounds even if dark nucleons carry baryon number \cite{McKeen:2018xwc}.} The dark shower events are generated using the Hidden Valley Module~\cite{Carloni:2010tw,Carloni:2011kk} of Pythia~\cite{Bierlich:2022pfr}. The neutral dark pion multiplicity distributions are shown in Fig.~\ref{fig:mul} with the dark pion mass fixed to 100 MeV. As shown in Fig.~\ref{fig:mul}, the neutral dark pion multiplicity increases for larger scalar masses. 
This is expected, since a heavier scalar mediator provides a larger energy budget for the dark shower, allowing more dark hadrons to be produced in the final state. 
For a fixed scalar mass, the multiplicity is also higher for smaller $m_{\rho_D}$. 
A lighter $\rho_D$ corresponds to a lower dark confinement scale $\Lambda_D$, as indicated in Eq.~(\ref{eq:mass}). This increases the available shower evolution range, allowing more splittings to occur before hadronization. 
As a result, benchmarks with smaller $m_{\rho_D}$ tend to produce more neutral dark pions. We have checked that the neutral dark pion multiplicity is rather insensitive to the dark quark mass with fixed $m_{\pi_D}$ and $m_{\rho_D}$.

The neutral pion can mix with the axion-like state $a$ when dark isospin is broken. The dark-isospin singlet $\eta'_D$ state is expected to be heavier due to anomaly effects and can be integrated out in a low-energy description. The mass term in the low-energy chiral Lagrangian is 
\beq
\mathcal{L}_\chi = \frac{f_D^2B_D}{2}\text{Tr}\left[(M_\chi + i a G_\chi)\Sigma_D + \text{h.c.}\right],
\eeq
where the dark quark mass matrix $M_\chi = \text{diag}\{m_{\chi_1},m_{\chi_2}\}$, $G_\chi = \text{diag}\{g_{\chi_1},g_{\chi_2}\}$, $\Sigma_D = \text{exp}(i\Pi_D/f_D)$, and $\Pi_D = \pi_D^a\tau^a$; here $\tau^a$ are Pauli matrices.
The low-energy constant $B_D$ determines the neutral dark pion mass through the relation $m_{\pi_D}^2 = B_D(m_{\chi_1}+m_{\chi_2})$.
The neutral dark pion and $a$ mix through the dark-quark mass matrix and the pseudoscalar couplings
\beq
\mathcal{L}_{\rm mix} = -\mu^2_{a\pi^0_D}a\pi^0_D,
\eeq
where $\mu^2_{a\pi^0_D} = B_Df_D(g_{a\chi_1}-g_{a\chi_2})$.  For a small mixing angle
\beq
\theta_{a\pi^0_D} \simeq \frac{\mu^2_{a\pi^0_D}}{m_a^2-m^2_{\pi_D}}.
\eeq
The physical neutral dark pion inherits the
coupling of $a$ to electrons,
\begin{equation}
  g_{\pi^0_D e}\simeq \theta_{a\pi^0_D} g_{ae} \simeq  \frac{\Delta m_e\, f_D m^2_{\pi_D}}{v_\Phi^2(m_a^2 - m^2_{\pi_D})}\frac{m_{\chi_1}-m_{\chi_2}}{m_{\chi_1}+m_{\chi_2}}.
\end{equation}
This induces the decay $\pi_D^0 \to e^+ e^-$.
The decay width can be written as
\begin{equation}
  \Gamma(\pi_D^0 \to e^+e^-)=\frac{g_{\pi^0_D e}^2}{8\pi} m_{\pi_D}
  \sqrt{1 - \frac{4m_e^2}{m_{\pi_D}^2}}.
\end{equation}
We fix $m_a = 15$~MeV and $m_{\chi_1} = 3 m_{\chi_2}$. In the case of $m_{\pi_D}\gg m_a$, the proper decay length of dark neutral pion is 
\bea\nonumber
l_{\pi_D}&\simeq& 0.25~\text{m}\left(\frac{f_D}{0.1~\text{GeV}}\right)^{-2}\left(\frac{m_{\pi_D}}{0.3~\text{GeV}}\right)^{-1}
\\
&\times&
\left(
\frac{\Delta m_e}{m_e}\right)^2\left(\frac{g_{Se}}{5.0\times 10^{-5}}\right)^{-4},
\label{eq:lpiD}
\eea
where we used the relations $v_\Phi = \Delta m_e/g_{Se}$ and $g_{ae} = g_{Se}$.

\begin{figure}[t]
	\centering
\includegraphics[width=\columnwidth]{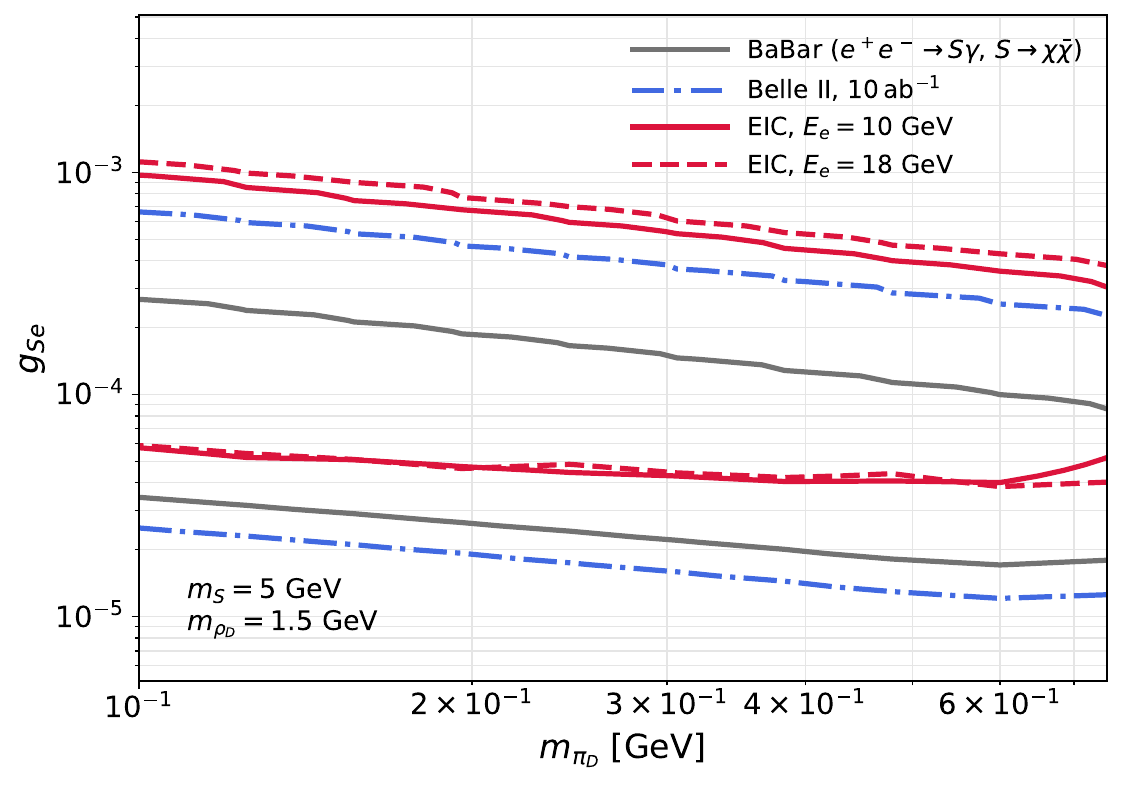}
    \includegraphics[width=\columnwidth]{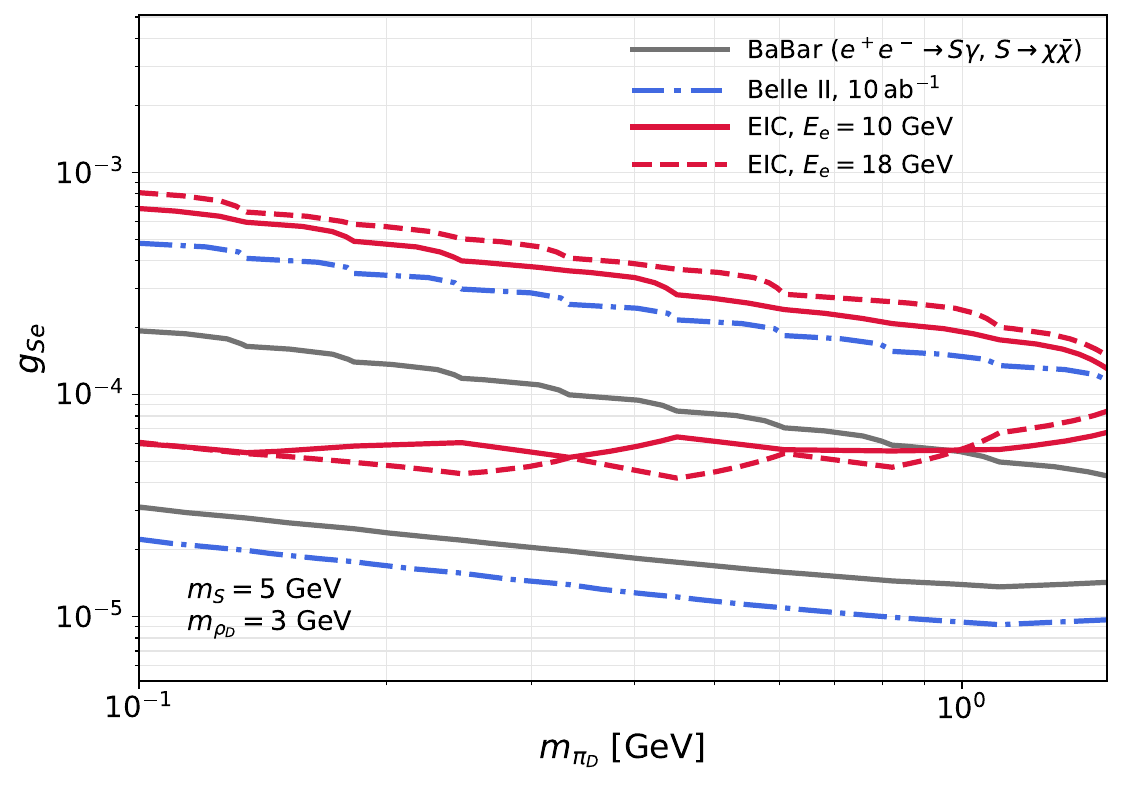}
\caption{The 3-event contours for $m_S$ = 5 GeV;  $\rho_D$ mass is fixed to be 1.5 (3.0) GeV in the top (bottom) panel.  The solid (dashed) red curves corresponds to projections with an electron beam energy of 10 (18) GeV.  Limits corresponding to the BaBar integrated luminosity of 514 \invfb are provided by the solid black contour.  The projected reach of Belle II, assuming an integrated luminosity of 10  ab$^{-1}$, are presented by the dotted dash blue contours.  For each of the experiments, upper limits roughly correspond to the resolution for short-distance tracks, while lower limits are set by the production rate for $S$.}
\label{fig:contour5GeV}
\end{figure}

\begin{figure}[t]
	\centering
\includegraphics[width=\columnwidth]{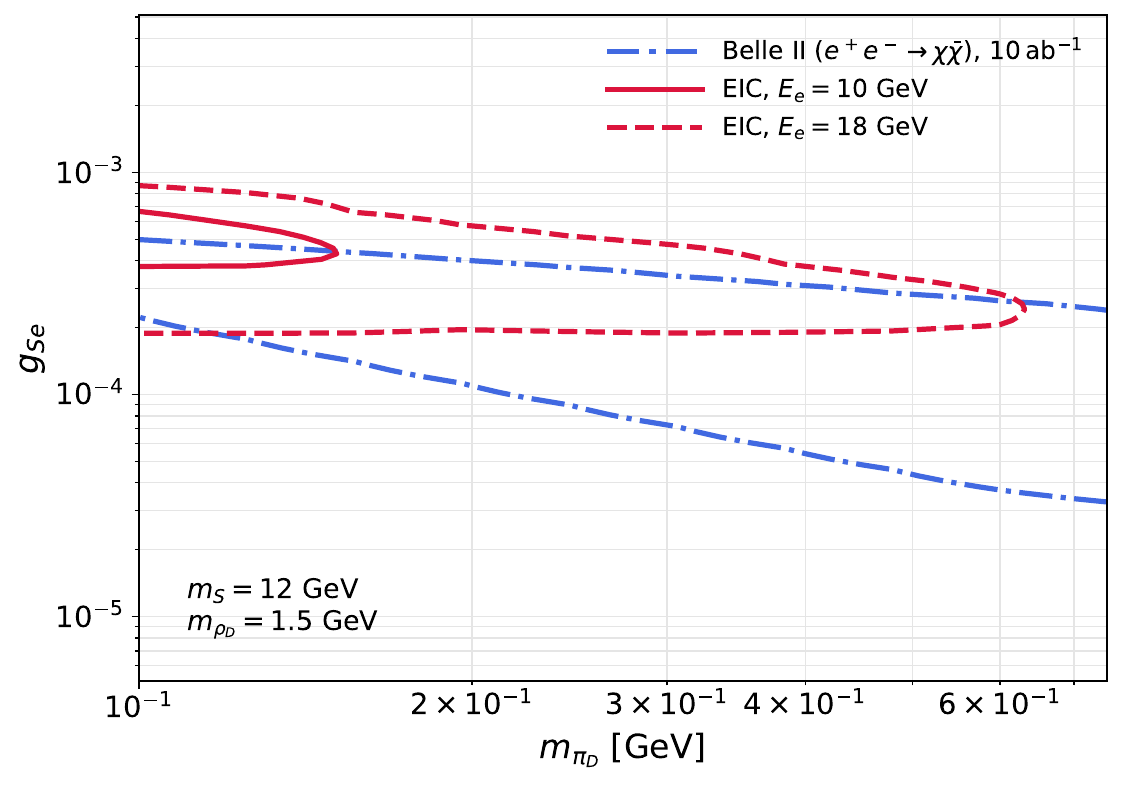}
    \includegraphics[width=\columnwidth]{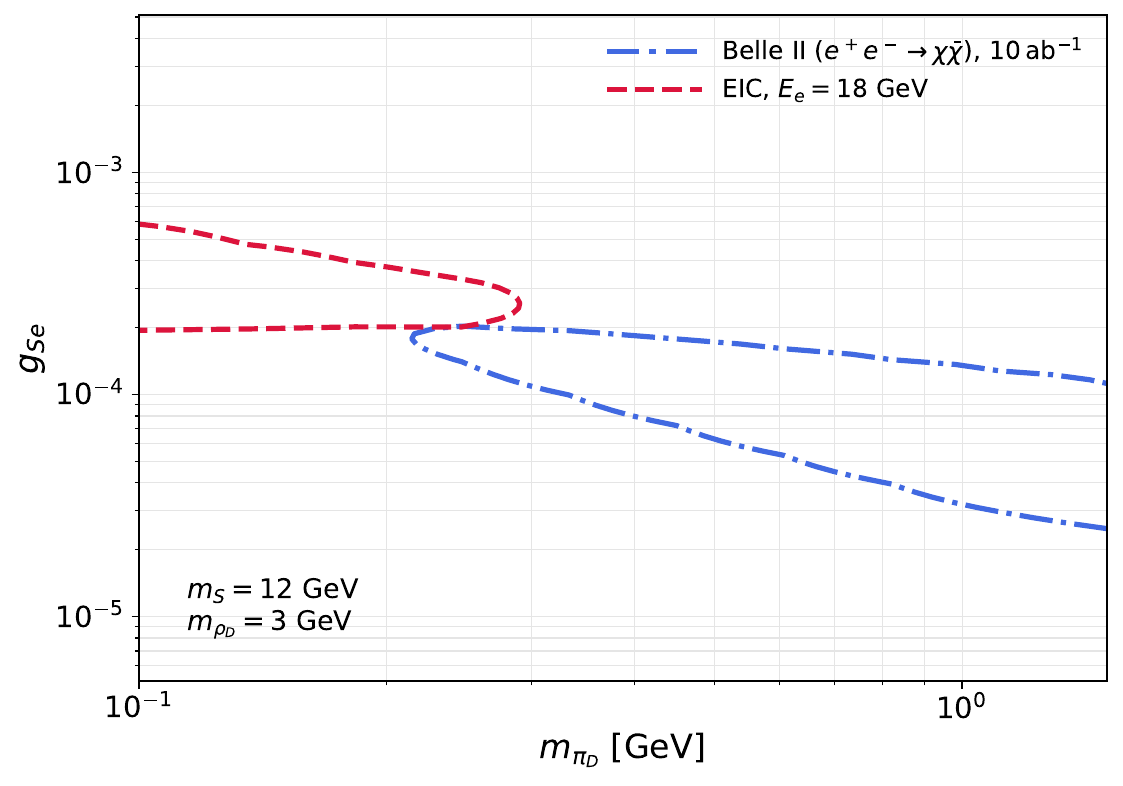}
	\caption{The same as  Fig.~\ref{fig:contour5GeV} but for $m_S$ = 12 GeV.  Note that Belle II can only produce dark quark pairs through off-shell $S$ via $e^- e^+ \to\chi\bar\chi$, since the mediator mass $m_S$ is outside their kinematic reach.  As a result, the Belle-II limits cut off at small $m_{\pi_D}$ as the dark quark Yukawa coupling becomes smaller.}
	\label{fig:contour10GeV}
\end{figure}

The charged dark pions, absent any flavor-changing interactions in that sector, are rendered stable and will lead to missing energy signals.  We do not expect these states to lead to unwanted cosmological effects for the benchmark parameters considered, as they are expected to annihilate efficiently into the unstable neutral dark pions and end up with a tiny relic abundance.  We will not discuss the potential cosmological implications of our setup further, as they are outside the scope of the present work.

The EIC is especially attractive for this search because the visible decay
products from GeV-scale dark pions can be soft. A detector and data acquisition
strategy that preserves soft particles can therefore probe regions of parameter
space that are difficult for conventional high-threshold triggers~\cite{Adkins:2022jfp}. 
The benchmark signal requirement is that at least two neutral dark pions decay within a fiducial volume,
\begin{equation}
  L_{\rm min} < L_{\pi_D} < L_{\rm max},
\end{equation}
where a representative EIC range setting is $L_{\rm min}=100~\mu$m and $L_{\rm max}=1$~m  \cite{Davoudiasl:2023pkq}. Additional requirements must be imposed on the neutral dark pion energy and pseudorapidity to ensure a good reconstruction accuracy
\begin{equation}
  E_{\pi_D} > 0.5~{\rm GeV}, \qquad |\eta_{\pi_D}| < 3.5 .
\end{equation}

 We set $E_e = 10~\rm{or}~18$~GeV  and $E_A = 100$~GeV/nucleon. We use gold with $Z=79$ and $A=197$ as the heavy ion. The ion form factor is adopted from~\cite{Helm:1956zz}.  For a given dark pion with momentum $p_{\pi_D}$, the mean lab-frame decay length is 
\begin{equation}
  \bar{L} =
  \frac{p_{\pi_D}}{m_{\pi_D}}l_{\pi_D} .
\end{equation}
The probability that this particle decays inside a fiducial interval is
\begin{equation}
  P_{\rm fid}
  =
  \exp\left(-\frac{L_{\rm min}}{\bar{L}}\right)
  -
  \exp\left(-\frac{L_{\rm max}}{\bar{L}}\right).
\end{equation}

We select signal events with at least two valid displaced decays, and assume a veto on incoherent scattering of the nucleus using the zero-degree calorimeter.  Furthermore, significant missing energy due to the production of stable $\pi_D^{\pm}$ can be used as an additional handle to remove the SM background as studied in~\cite{Davoudiasl:2025rpn}.  Hence, we expect this search to be nearly background free, due to the aforementioned distinct features of the signal. 
Then, the expected number of signal events is
\begin{equation}
  N_S =
  \sigma(eA\to eAS)\,
  {\cal L}\,P(\geq 2~{\rm fiducial~decays}),
\end{equation}
where $\cal L = $100/$A$ \invfb is the assumed integrated luminosity. We consider two benchmark scenarios with $m_S = 5~\mathrm{GeV}$ and $12~\mathrm{GeV}$, respectively. The $N_S = 3$ contours are shown in Figs.~\ref{fig:contour5GeV} and \ref{fig:contour10GeV}. At $m_S=5$~GeV, we also include the current bounds from BaBar using the data of 514 \invfb. For the analysis of BaBar, we require~\cite{BaBar:2001yhh,BaBar:2015jvu}
\beq
1~\text{cm}<r_{\perp}< 50~\text{cm}, -1.0<\eta_{\pi_D}< 1.9, E_{\pi_D}>0.5~\text{GeV}.
\eeq
Similarly, we adopt the following cuts for Belle II with 10 ab$^{-1}$ integrated luminosity~\cite{Belle-II:2023ueh,Jaeckel:2023huy}
\beq
0.05~\text{cm}<r_{\perp}< 60~\text{cm}, -1.32<\eta_{\pi_D}< 1.25, E_{\pi_D}>0.5~\text{GeV}.
\eeq
As we can see from Fig.~\ref{fig:contour5GeV}, for a light mediator $S$, within the kinematic reach of BaBar and Belle II, the large data samples allow probing $g_{Se}$ couplings below the projections for the EIC.  However, the expected vertex resolution of the EIC  can allow a better reach for larger couplings.  This is a feature of our model with $g_{Se}=g_{ae}$ since larger scalar couplings coincide with larger couplings to $a$,  and hence less displaced dark pion vertices.

In Fig.~\ref{fig:contour10GeV}, we have chosen $m_S$ outside the reach of the $B$ meson factories.  Thus, the dark quark $\chi_i$ production would go through an off-shell mediator at those facilities and it would hence depend on the Yukawa coupling of $\chi_i$ which gets smaller as dark pion masses $m_{\pi_D}$ get smaller.  However, the larger kinematic reach of the EIC allows on-shell production of the mediator and it remains insensitive to small dark quark Yukawa couplings, giving it an advantage at low $m_{\pi_D}$.  As a result, the EIC and Belle-II searches are complementary to one another in this part of our parameter space.

\section{Summary} 

A new confining sector can arise in a variety of models that address open fundamental questions.  In this work, we considered a ``dark QCD'' model whose confinement scale is in the GeV regime, which may be motivated in the context of an asymmetric dark matter scenario, for example.  We postulated that this sector couples to the SM, and in particular to electrons, via feeble interactions mediated by a new complex scalar.  The condensation of this scalar can provide masses for dark quarks and its associated Goldstone boson allows flavor diagonal ``dark pions" to decay into electrons, through mixing.  Flavor off-diagonal dark pions would typically be stable and escape the detector.      

Under the above assumptions, we showed that the EIC can uncover dark hadrons, using distinct displaced vertex and missing energy signals.    
We examined the potential of the EIC to explore the dark QCD phenomenology, assuming both an initial electron beam energy of 10 GeV and a possible upgrade to 18 GeV.  Scalar mediator masses within and beyond the kinematic reach of the Belle II experiment, which can probe some of the same physics, were also considered.  

Our results suggest that in the cases studied here, the EIC can in principle cover interesting new regions of the parameter space of the model which will otherwise remain unexplored in the coming years.  While we only considered a dark analog of the SM strong interactions, we expect that a similar conclusion would hold for other theories whose confinement scales are in the same $\ord{\rm GeV}$ regime and couple to the SM near the levels considered in our work.  One may also consider alternative mediators for producing the dark quarks, such as dark photons, which would lead to a different phenomenology but qualitatively similar conclusions. Hence, the EIC can be a venue for investigating not only ordinary nuclear matter, but also possible dark hadronic constituents of the cosmos, providing a better understanding of the visible and invisible worlds.          
        
\vskip0.5cm
The digital data related to this work can be found as
ancillary files with the arXiv submission.

	~
	\begin{acknowledgments}
		The work of H.D. and H.L. is supported by the US Department of Energy under Grant Contract DE-SC0012704.  The work of E.T.N. is supported by the US Department of Energy under Grant Contract DE-SC0010005.

        Code development was performed with assistance from OpenAI's ChatGPT and Codex under human supervision.  The authors take full responsibility for the physics content of this work.
 	\end{acknowledgments}

        \bibliography{ref.bib}

\begin{thebibliography}{46}%
\makeatletter
\providecommand \@ifxundefined [1]{%
 \@ifx{#1\undefined}
}%
\providecommand \@ifnum [1]{%
 \ifnum #1\expandafter \@firstoftwo
 \else \expandafter \@secondoftwo
 \fi
}%
\providecommand \@ifx [1]{%
 \ifx #1\expandafter \@firstoftwo
 \else \expandafter \@secondoftwo
 \fi
}%
\providecommand \natexlab [1]{#1}%
\providecommand \enquote  [1]{``#1''}%
\providecommand \bibnamefont  [1]{#1}%
\providecommand \bibfnamefont [1]{#1}%
\providecommand \citenamefont [1]{#1}%
\providecommand \href@noop [0]{\@secondoftwo}%
\providecommand \href [0]{\begingroup \@sanitize@url \@href}%
\providecommand \@href[1]{\@@startlink{#1}\@@href}%
\providecommand \@@href[1]{\endgroup#1\@@endlink}%
\providecommand \@sanitize@url [0]{\catcode `\\12\catcode `\$12\catcode
  `\&12\catcode `\#12\catcode `\^12\catcode `\_12\catcode `\%12\relax}%
\providecommand \@@startlink[1]{}%
\providecommand \@@endlink[0]{}%
\providecommand \url  [0]{\begingroup\@sanitize@url \@url }%
\providecommand \@url [1]{\endgroup\@href {#1}{\urlprefix }}%
\providecommand \urlprefix  [0]{URL }%
\providecommand \Eprint [0]{\href }%
\providecommand \doibase [0]{https://doi.org/}%
\providecommand \selectlanguage [0]{\@gobble}%
\providecommand \bibinfo  [0]{\@secondoftwo}%
\providecommand \bibfield  [0]{\@secondoftwo}%
\providecommand \translation [1]{[#1]}%
\providecommand \BibitemOpen [0]{}%
\providecommand \bibitemStop [0]{}%
\providecommand \bibitemNoStop [0]{.\EOS\space}%
\providecommand \EOS [0]{\spacefactor3000\relax}%
\providecommand \BibitemShut  [1]{\csname bibitem#1\endcsname}%
\let\auto@bib@innerbib\@empty
\bibitem [{\citenamefont {Cline}(2022)}]{Cline:2021itd}%
  \BibitemOpen
  \bibfield  {author} {\bibinfo {author} {\bibfnamefont {J.~M.}\ \bibnamefont
  {Cline}},\ }\bibfield  {title} {\bibinfo {title} {{Dark atoms and composite
  dark matter}},\ }\href {https://doi.org/10.21468/SciPostPhysLectNotes.52}
  {\bibfield  {journal} {\bibinfo  {journal} {SciPost Phys. Lect. Notes}\
  }\textbf {\bibinfo {volume} {52}},\ \bibinfo {pages} {1} (\bibinfo {year}
  {2022})},\ \Eprint {https://arxiv.org/abs/2108.10314} {arXiv:2108.10314
  [hep-ph]} \BibitemShut {NoStop}%
\bibitem [{\citenamefont {Asadi}\ \emph {et~al.}(2026)\citenamefont {Asadi},
  \citenamefont {Batz},\ and\ \citenamefont {Kribs}}]{Asadi:2026mip}%
  \BibitemOpen
  \bibfield  {author} {\bibinfo {author} {\bibfnamefont {P.}~\bibnamefont
  {Asadi}}, \bibinfo {author} {\bibfnamefont {A.}~\bibnamefont {Batz}},\ and\
  \bibinfo {author} {\bibfnamefont {G.~D.}\ \bibnamefont {Kribs}},\ }\bibfield
  {title} {\bibinfo {title} {{Rich Phenomenology from Simple Ingredients: A
  Review of Confining Dark Sectors}},\ }\href@noop {} {\  (\bibinfo {year}
  {2026})},\ \Eprint {https://arxiv.org/abs/2606.30760} {arXiv:2606.30760
  [hep-ph]} \BibitemShut {NoStop}%
\bibitem [{\citenamefont {Davoudiasl}\ and\ \citenamefont
  {Mohapatra}(2012)}]{Davoudiasl:2012uw}%
  \BibitemOpen
  \bibfield  {author} {\bibinfo {author} {\bibfnamefont {H.}~\bibnamefont
  {Davoudiasl}}\ and\ \bibinfo {author} {\bibfnamefont {R.~N.}\ \bibnamefont
  {Mohapatra}},\ }\bibfield  {title} {\bibinfo {title} {{On Relating the
  Genesis of Cosmic Baryons and Dark Matter}},\ }\href
  {https://doi.org/10.1088/1367-2630/14/9/095011} {\bibfield  {journal}
  {\bibinfo  {journal} {New J. Phys.}\ }\textbf {\bibinfo {volume} {14}},\
  \bibinfo {pages} {095011} (\bibinfo {year} {2012})},\ \Eprint
  {https://arxiv.org/abs/1203.1247} {arXiv:1203.1247 [hep-ph]} \BibitemShut
  {NoStop}%
\bibitem [{\citenamefont {Petraki}\ and\ \citenamefont
  {Volkas}(2013)}]{Petraki:2013wwa}%
  \BibitemOpen
  \bibfield  {author} {\bibinfo {author} {\bibfnamefont {K.}~\bibnamefont
  {Petraki}}\ and\ \bibinfo {author} {\bibfnamefont {R.~R.}\ \bibnamefont
  {Volkas}},\ }\bibfield  {title} {\bibinfo {title} {{Review of asymmetric dark
  matter}},\ }\href {https://doi.org/10.1142/S0217751X13300287} {\bibfield
  {journal} {\bibinfo  {journal} {Int. J. Mod. Phys. A}\ }\textbf {\bibinfo
  {volume} {28}},\ \bibinfo {pages} {1330028} (\bibinfo {year} {2013})},\
  \Eprint {https://arxiv.org/abs/1305.4939} {arXiv:1305.4939 [hep-ph]}
  \BibitemShut {NoStop}%
\bibitem [{\citenamefont {Zurek}(2014)}]{Zurek:2013wia}%
  \BibitemOpen
  \bibfield  {author} {\bibinfo {author} {\bibfnamefont {K.~M.}\ \bibnamefont
  {Zurek}},\ }\bibfield  {title} {\bibinfo {title} {{Asymmetric Dark Matter:
  Theories, Signatures, and Constraints}},\ }\href
  {https://doi.org/10.1016/j.physrep.2013.12.001} {\bibfield  {journal}
  {\bibinfo  {journal} {Phys. Rept.}\ }\textbf {\bibinfo {volume} {537}},\
  \bibinfo {pages} {91} (\bibinfo {year} {2014})},\ \Eprint
  {https://arxiv.org/abs/1308.0338} {arXiv:1308.0338 [hep-ph]} \BibitemShut
  {NoStop}%
\bibitem [{\citenamefont {Han}\ \emph {et~al.}(2008)\citenamefont {Han},
  \citenamefont {Si}, \citenamefont {Zurek},\ and\ \citenamefont
  {Strassler}}]{Han:2007ae}%
  \BibitemOpen
  \bibfield  {author} {\bibinfo {author} {\bibfnamefont {T.}~\bibnamefont
  {Han}}, \bibinfo {author} {\bibfnamefont {Z.}~\bibnamefont {Si}}, \bibinfo
  {author} {\bibfnamefont {K.~M.}\ \bibnamefont {Zurek}},\ and\ \bibinfo
  {author} {\bibfnamefont {M.~J.}\ \bibnamefont {Strassler}},\ }\bibfield
  {title} {\bibinfo {title} {{Phenomenology of hidden valleys at hadron
  colliders}},\ }\href {https://doi.org/10.1088/1126-6708/2008/07/008}
  {\bibfield  {journal} {\bibinfo  {journal} {JHEP}\ }\textbf {\bibinfo
  {volume} {07}},\ \bibinfo {pages} {008}},\ \Eprint
  {https://arxiv.org/abs/0712.2041} {arXiv:0712.2041 [hep-ph]} \BibitemShut
  {NoStop}%
\bibitem [{\citenamefont {Baumgart}\ \emph {et~al.}(2009)\citenamefont
  {Baumgart}, \citenamefont {Cheung}, \citenamefont {Ruderman}, \citenamefont
  {Wang},\ and\ \citenamefont {Yavin}}]{Baumgart:2009tn}%
  \BibitemOpen
  \bibfield  {author} {\bibinfo {author} {\bibfnamefont {M.}~\bibnamefont
  {Baumgart}}, \bibinfo {author} {\bibfnamefont {C.}~\bibnamefont {Cheung}},
  \bibinfo {author} {\bibfnamefont {J.~T.}\ \bibnamefont {Ruderman}}, \bibinfo
  {author} {\bibfnamefont {L.-T.}\ \bibnamefont {Wang}},\ and\ \bibinfo
  {author} {\bibfnamefont {I.}~\bibnamefont {Yavin}},\ }\bibfield  {title}
  {\bibinfo {title} {{Non-Abelian Dark Sectors and Their Collider
  Signatures}},\ }\href {https://doi.org/10.1088/1126-6708/2009/04/014}
  {\bibfield  {journal} {\bibinfo  {journal} {JHEP}\ }\textbf {\bibinfo
  {volume} {04}},\ \bibinfo {pages} {014}},\ \Eprint
  {https://arxiv.org/abs/0901.0283} {arXiv:0901.0283 [hep-ph]} \BibitemShut
  {NoStop}%
\bibitem [{\citenamefont {Falkowski}\ \emph {et~al.}(2010)\citenamefont
  {Falkowski}, \citenamefont {Ruderman}, \citenamefont {Volansky},\ and\
  \citenamefont {Zupan}}]{Falkowski:2010cm}%
  \BibitemOpen
  \bibfield  {author} {\bibinfo {author} {\bibfnamefont {A.}~\bibnamefont
  {Falkowski}}, \bibinfo {author} {\bibfnamefont {J.~T.}\ \bibnamefont
  {Ruderman}}, \bibinfo {author} {\bibfnamefont {T.}~\bibnamefont {Volansky}},\
  and\ \bibinfo {author} {\bibfnamefont {J.}~\bibnamefont {Zupan}},\ }\bibfield
   {title} {\bibinfo {title} {{Hidden Higgs Decaying to Lepton Jets}},\ }\href
  {https://doi.org/10.1007/JHEP05(2010)077} {\bibfield  {journal} {\bibinfo
  {journal} {JHEP}\ }\textbf {\bibinfo {volume} {05}},\ \bibinfo {pages}
  {077}},\ \Eprint {https://arxiv.org/abs/1002.2952} {arXiv:1002.2952 [hep-ph]}
  \BibitemShut {NoStop}%
\bibitem [{\citenamefont {Cohen}\ \emph {et~al.}(2015)\citenamefont {Cohen},
  \citenamefont {Lisanti},\ and\ \citenamefont {Lou}}]{Cohen:2015toa}%
  \BibitemOpen
  \bibfield  {author} {\bibinfo {author} {\bibfnamefont {T.}~\bibnamefont
  {Cohen}}, \bibinfo {author} {\bibfnamefont {M.}~\bibnamefont {Lisanti}},\
  and\ \bibinfo {author} {\bibfnamefont {H.~K.}\ \bibnamefont {Lou}},\
  }\bibfield  {title} {\bibinfo {title} {{Semivisible Jets: Dark Matter
  Undercover at the LHC}},\ }\href
  {https://doi.org/10.1103/PhysRevLett.115.171804} {\bibfield  {journal}
  {\bibinfo  {journal} {Phys. Rev. Lett.}\ }\textbf {\bibinfo {volume} {115}},\
  \bibinfo {pages} {171804} (\bibinfo {year} {2015})},\ \Eprint
  {https://arxiv.org/abs/1503.00009} {arXiv:1503.00009 [hep-ph]} \BibitemShut
  {NoStop}%
\bibitem [{\citenamefont {Schwaller}\ \emph {et~al.}(2015)\citenamefont
  {Schwaller}, \citenamefont {Stolarski},\ and\ \citenamefont
  {Weiler}}]{Schwaller:2015gea}%
  \BibitemOpen
  \bibfield  {author} {\bibinfo {author} {\bibfnamefont {P.}~\bibnamefont
  {Schwaller}}, \bibinfo {author} {\bibfnamefont {D.}~\bibnamefont
  {Stolarski}},\ and\ \bibinfo {author} {\bibfnamefont {A.}~\bibnamefont
  {Weiler}},\ }\bibfield  {title} {\bibinfo {title} {{Emerging Jets}},\ }\href
  {https://doi.org/10.1007/JHEP05(2015)059} {\bibfield  {journal} {\bibinfo
  {journal} {JHEP}\ }\textbf {\bibinfo {volume} {05}},\ \bibinfo {pages}
  {059}},\ \Eprint {https://arxiv.org/abs/1502.05409} {arXiv:1502.05409
  [hep-ph]} \BibitemShut {NoStop}%
\bibitem [{\citenamefont {Cohen}\ \emph {et~al.}(2017)\citenamefont {Cohen},
  \citenamefont {Lisanti}, \citenamefont {Lou},\ and\ \citenamefont
  {Mishra-Sharma}}]{Cohen:2017pzm}%
  \BibitemOpen
  \bibfield  {author} {\bibinfo {author} {\bibfnamefont {T.}~\bibnamefont
  {Cohen}}, \bibinfo {author} {\bibfnamefont {M.}~\bibnamefont {Lisanti}},
  \bibinfo {author} {\bibfnamefont {H.~K.}\ \bibnamefont {Lou}},\ and\ \bibinfo
  {author} {\bibfnamefont {S.}~\bibnamefont {Mishra-Sharma}},\ }\bibfield
  {title} {\bibinfo {title} {{LHC Searches for Dark Sector Showers}},\ }\href
  {https://doi.org/10.1007/JHEP11(2017)196} {\bibfield  {journal} {\bibinfo
  {journal} {JHEP}\ }\textbf {\bibinfo {volume} {11}},\ \bibinfo {pages}
  {196}},\ \Eprint {https://arxiv.org/abs/1707.05326} {arXiv:1707.05326
  [hep-ph]} \BibitemShut {NoStop}%
\bibitem [{\citenamefont {Knapen}\ \emph {et~al.}(2021)\citenamefont {Knapen},
  \citenamefont {Shelton},\ and\ \citenamefont {Xu}}]{Knapen:2021eip}%
  \BibitemOpen
  \bibfield  {author} {\bibinfo {author} {\bibfnamefont {S.}~\bibnamefont
  {Knapen}}, \bibinfo {author} {\bibfnamefont {J.}~\bibnamefont {Shelton}},\
  and\ \bibinfo {author} {\bibfnamefont {D.}~\bibnamefont {Xu}},\ }\bibfield
  {title} {\bibinfo {title} {{Perturbative benchmark models for a dark shower
  search program}},\ }\href {https://doi.org/10.1103/PhysRevD.103.115013}
  {\bibfield  {journal} {\bibinfo  {journal} {Phys. Rev. D}\ }\textbf {\bibinfo
  {volume} {103}},\ \bibinfo {pages} {115013} (\bibinfo {year} {2021})},\
  \Eprint {https://arxiv.org/abs/2103.01238} {arXiv:2103.01238 [hep-ph]}
  \BibitemShut {NoStop}%
\bibitem [{\citenamefont {Albouy}\ \emph {et~al.}(2022)\citenamefont {Albouy}
  \emph {et~al.}}]{Albouy:2022cin}%
  \BibitemOpen
  \bibfield  {author} {\bibinfo {author} {\bibfnamefont {G.}~\bibnamefont
  {Albouy}} \emph {et~al.},\ }\bibfield  {title} {\bibinfo {title} {{Theory,
  phenomenology, and experimental avenues for dark showers: a Snowmass 2021
  report}},\ }\href {https://doi.org/10.1140/epjc/s10052-022-11048-8}
  {\bibfield  {journal} {\bibinfo  {journal} {Eur. Phys. J. C}\ }\textbf
  {\bibinfo {volume} {82}},\ \bibinfo {pages} {1132} (\bibinfo {year}
  {2022})},\ \Eprint {https://arxiv.org/abs/2203.09503} {arXiv:2203.09503
  [hep-ph]} \BibitemShut {NoStop}%
\bibitem [{\citenamefont {Cohen}\ \emph {et~al.}(2023)\citenamefont {Cohen},
  \citenamefont {Roloff},\ and\ \citenamefont {Scherb}}]{Cohen:2023mya}%
  \BibitemOpen
  \bibfield  {author} {\bibinfo {author} {\bibfnamefont {T.}~\bibnamefont
  {Cohen}}, \bibinfo {author} {\bibfnamefont {J.}~\bibnamefont {Roloff}},\ and\
  \bibinfo {author} {\bibfnamefont {C.}~\bibnamefont {Scherb}},\ }\bibfield
  {title} {\bibinfo {title} {{Dark sector showers in the Lund jet plane}},\
  }\href {https://doi.org/10.1103/PhysRevD.108.L031501} {\bibfield  {journal}
  {\bibinfo  {journal} {Phys. Rev. D}\ }\textbf {\bibinfo {volume} {108}},\
  \bibinfo {pages} {L031501} (\bibinfo {year} {2023})},\ \Eprint
  {https://arxiv.org/abs/2301.07732} {arXiv:2301.07732 [hep-ph]} \BibitemShut
  {NoStop}%
\bibitem [{\citenamefont {Cheng}\ \emph {et~al.}(2024)\citenamefont {Cheng},
  \citenamefont {Jiang}, \citenamefont {Li},\ and\ \citenamefont
  {Salvioni}}]{Cheng:2024hvq}%
  \BibitemOpen
  \bibfield  {author} {\bibinfo {author} {\bibfnamefont {H.-C.}\ \bibnamefont
  {Cheng}}, \bibinfo {author} {\bibfnamefont {X.-H.}\ \bibnamefont {Jiang}},
  \bibinfo {author} {\bibfnamefont {L.}~\bibnamefont {Li}},\ and\ \bibinfo
  {author} {\bibfnamefont {E.}~\bibnamefont {Salvioni}},\ }\bibfield  {title}
  {\bibinfo {title} {{Dark showers from Z-dark Z' mixing}},\ }\href
  {https://doi.org/10.1007/JHEP04(2024)081} {\bibfield  {journal} {\bibinfo
  {journal} {JHEP}\ }\textbf {\bibinfo {volume} {04}},\ \bibinfo {pages}
  {081}},\ \Eprint {https://arxiv.org/abs/2401.08785} {arXiv:2401.08785
  [hep-ph]} \BibitemShut {NoStop}%
\bibitem [{\citenamefont {Cheng}\ \emph {et~al.}(2025)\citenamefont {Cheng},
  \citenamefont {Jiang},\ and\ \citenamefont {Li}}]{Cheng:2024aco}%
  \BibitemOpen
  \bibfield  {author} {\bibinfo {author} {\bibfnamefont {H.-C.}\ \bibnamefont
  {Cheng}}, \bibinfo {author} {\bibfnamefont {X.-H.}\ \bibnamefont {Jiang}},\
  and\ \bibinfo {author} {\bibfnamefont {L.}~\bibnamefont {Li}},\ }\bibfield
  {title} {\bibinfo {title} {{Phenomenology of electroweak portal dark showers:
  high energy direct probes and low energy complementarity}},\ }\href
  {https://doi.org/10.1007/JHEP01(2025)149} {\bibfield  {journal} {\bibinfo
  {journal} {JHEP}\ }\textbf {\bibinfo {volume} {01}},\ \bibinfo {pages}
  {149}},\ \Eprint {https://arxiv.org/abs/2408.13304} {arXiv:2408.13304
  [hep-ph]} \BibitemShut {NoStop}%
\bibitem [{\citenamefont {Bernreuther}\ \emph {et~al.}(2022)\citenamefont
  {Bernreuther}, \citenamefont {B{\"o}se}, \citenamefont {Ferber},
  \citenamefont {Hearty}, \citenamefont {Kahlhoefer}, \citenamefont
  {Morandini},\ and\ \citenamefont {Schmidt-Hoberg}}]{Bernreuther:2022jlj}%
  \BibitemOpen
  \bibfield  {author} {\bibinfo {author} {\bibfnamefont {E.}~\bibnamefont
  {Bernreuther}}, \bibinfo {author} {\bibfnamefont {K.}~\bibnamefont
  {B{\"o}se}}, \bibinfo {author} {\bibfnamefont {T.}~\bibnamefont {Ferber}},
  \bibinfo {author} {\bibfnamefont {C.}~\bibnamefont {Hearty}}, \bibinfo
  {author} {\bibfnamefont {F.}~\bibnamefont {Kahlhoefer}}, \bibinfo {author}
  {\bibfnamefont {A.}~\bibnamefont {Morandini}},\ and\ \bibinfo {author}
  {\bibfnamefont {K.}~\bibnamefont {Schmidt-Hoberg}},\ }\bibfield  {title}
  {\bibinfo {title} {{Forecasting dark showers at Belle II}},\ }\href
  {https://doi.org/10.1007/JHEP12(2022)005} {\bibfield  {journal} {\bibinfo
  {journal} {JHEP}\ }\textbf {\bibinfo {volume} {12}},\ \bibinfo {pages}
  {005}},\ \Eprint {https://arxiv.org/abs/2203.08824} {arXiv:2203.08824
  [hep-ph]} \BibitemShut {NoStop}%
\bibitem [{\citenamefont {Lu}\ and\ \citenamefont {Xi}(2025)}]{Lu:2025cty}%
  \BibitemOpen
  \bibfield  {author} {\bibinfo {author} {\bibfnamefont {C.-T.}\ \bibnamefont
  {Lu}}\ and\ \bibinfo {author} {\bibfnamefont {C.}~\bibnamefont {Xi}},\
  }\bibfield  {title} {\bibinfo {title} {{Searching for leptophilic composite
  asymmetric dark sector at e+e- colliders}},\ }\href
  {https://doi.org/10.1103/xsmv-f97b} {\bibfield  {journal} {\bibinfo
  {journal} {Phys. Rev. D}\ }\textbf {\bibinfo {volume} {112}},\ \bibinfo
  {pages} {095027} (\bibinfo {year} {2025})},\ \Eprint
  {https://arxiv.org/abs/2509.12504} {arXiv:2509.12504 [hep-ph]} \BibitemShut
  {NoStop}%
\bibitem [{\citenamefont {Adkins}\ \emph {et~al.}(2025)\citenamefont {Adkins}
  \emph {et~al.}}]{Adkins:2022jfp}%
  \BibitemOpen
  \bibfield  {author} {\bibinfo {author} {\bibfnamefont {J.~K.}\ \bibnamefont
  {Adkins}} \emph {et~al.},\ }\bibfield  {title} {\bibinfo {title} {{Design of
  the ECCE detector for the Electron Ion Collider}},\ }\href
  {https://doi.org/10.1016/j.nima.2025.170240} {\bibfield  {journal} {\bibinfo
  {journal} {Nucl. Instrum. Meth. A}\ }\textbf {\bibinfo {volume} {1073}},\
  \bibinfo {pages} {170240} (\bibinfo {year} {2025})},\ \Eprint
  {https://arxiv.org/abs/2209.02580} {arXiv:2209.02580 [physics.ins-det]}
  \BibitemShut {NoStop}%
\bibitem [{\citenamefont {Chacko}\ \emph
  {et~al.}(2006{\natexlab{a}})\citenamefont {Chacko}, \citenamefont {Goh},\
  and\ \citenamefont {Harnik}}]{Chacko:2005pe}%
  \BibitemOpen
  \bibfield  {author} {\bibinfo {author} {\bibfnamefont {Z.}~\bibnamefont
  {Chacko}}, \bibinfo {author} {\bibfnamefont {H.-S.}\ \bibnamefont {Goh}},\
  and\ \bibinfo {author} {\bibfnamefont {R.}~\bibnamefont {Harnik}},\
  }\bibfield  {title} {\bibinfo {title} {{The Twin Higgs: Natural electroweak
  breaking from mirror symmetry}},\ }\href
  {https://doi.org/10.1103/PhysRevLett.96.231802} {\bibfield  {journal}
  {\bibinfo  {journal} {Phys. Rev. Lett.}\ }\textbf {\bibinfo {volume} {96}},\
  \bibinfo {pages} {231802} (\bibinfo {year} {2006}{\natexlab{a}})},\ \Eprint
  {https://arxiv.org/abs/hep-ph/0506256} {arXiv:hep-ph/0506256} \BibitemShut
  {NoStop}%
\bibitem [{\citenamefont {Chacko}\ \emph
  {et~al.}(2006{\natexlab{b}})\citenamefont {Chacko}, \citenamefont {Goh},\
  and\ \citenamefont {Harnik}}]{Chacko:2005un}%
  \BibitemOpen
  \bibfield  {author} {\bibinfo {author} {\bibfnamefont {Z.}~\bibnamefont
  {Chacko}}, \bibinfo {author} {\bibfnamefont {H.-S.}\ \bibnamefont {Goh}},\
  and\ \bibinfo {author} {\bibfnamefont {R.}~\bibnamefont {Harnik}},\
  }\bibfield  {title} {\bibinfo {title} {{A Twin Higgs model from left-right
  symmetry}},\ }\href {https://doi.org/10.1088/1126-6708/2006/01/108}
  {\bibfield  {journal} {\bibinfo  {journal} {JHEP}\ }\textbf {\bibinfo
  {volume} {01}},\ \bibinfo {pages} {108}},\ \Eprint
  {https://arxiv.org/abs/hep-ph/0512088} {arXiv:hep-ph/0512088} \BibitemShut
  {NoStop}%
\bibitem [{\citenamefont {Falkowski}\ \emph {et~al.}(2006)\citenamefont
  {Falkowski}, \citenamefont {Pokorski},\ and\ \citenamefont
  {Schmaltz}}]{Falkowski:2006qq}%
  \BibitemOpen
  \bibfield  {author} {\bibinfo {author} {\bibfnamefont {A.}~\bibnamefont
  {Falkowski}}, \bibinfo {author} {\bibfnamefont {S.}~\bibnamefont
  {Pokorski}},\ and\ \bibinfo {author} {\bibfnamefont {M.}~\bibnamefont
  {Schmaltz}},\ }\bibfield  {title} {\bibinfo {title} {{Twin SUSY}},\ }\href
  {https://doi.org/10.1103/PhysRevD.74.035003} {\bibfield  {journal} {\bibinfo
  {journal} {Phys. Rev. D}\ }\textbf {\bibinfo {volume} {74}},\ \bibinfo
  {pages} {035003} (\bibinfo {year} {2006})},\ \Eprint
  {https://arxiv.org/abs/hep-ph/0604066} {arXiv:hep-ph/0604066} \BibitemShut
  {NoStop}%
\bibitem [{\citenamefont {Chang}\ \emph {et~al.}(2007)\citenamefont {Chang},
  \citenamefont {Hall},\ and\ \citenamefont {Weiner}}]{Chang:2006ra}%
  \BibitemOpen
  \bibfield  {author} {\bibinfo {author} {\bibfnamefont {S.}~\bibnamefont
  {Chang}}, \bibinfo {author} {\bibfnamefont {L.~J.}\ \bibnamefont {Hall}},\
  and\ \bibinfo {author} {\bibfnamefont {N.}~\bibnamefont {Weiner}},\
  }\bibfield  {title} {\bibinfo {title} {{A Supersymmetric twin Higgs}},\
  }\href {https://doi.org/10.1103/PhysRevD.75.035009} {\bibfield  {journal}
  {\bibinfo  {journal} {Phys. Rev. D}\ }\textbf {\bibinfo {volume} {75}},\
  \bibinfo {pages} {035009} (\bibinfo {year} {2007})},\ \Eprint
  {https://arxiv.org/abs/hep-ph/0604076} {arXiv:hep-ph/0604076} \BibitemShut
  {NoStop}%
\bibitem [{\citenamefont {Craig}\ \emph {et~al.}(2015)\citenamefont {Craig},
  \citenamefont {Knapen},\ and\ \citenamefont {Longhi}}]{Craig:2014aea}%
  \BibitemOpen
  \bibfield  {author} {\bibinfo {author} {\bibfnamefont {N.}~\bibnamefont
  {Craig}}, \bibinfo {author} {\bibfnamefont {S.}~\bibnamefont {Knapen}},\ and\
  \bibinfo {author} {\bibfnamefont {P.}~\bibnamefont {Longhi}},\ }\bibfield
  {title} {\bibinfo {title} {{Neutral Naturalness from Orbifold Higgs
  Models}},\ }\href {https://doi.org/10.1103/PhysRevLett.114.061803} {\bibfield
   {journal} {\bibinfo  {journal} {Phys. Rev. Lett.}\ }\textbf {\bibinfo
  {volume} {114}},\ \bibinfo {pages} {061803} (\bibinfo {year} {2015})},\
  \Eprint {https://arxiv.org/abs/1410.6808} {arXiv:1410.6808 [hep-ph]}
  \BibitemShut {NoStop}%
\bibitem [{\citenamefont {Barbieri}\ \emph {et~al.}(2015)\citenamefont
  {Barbieri}, \citenamefont {Greco}, \citenamefont {Rattazzi},\ and\
  \citenamefont {Wulzer}}]{Barbieri:2015lqa}%
  \BibitemOpen
  \bibfield  {author} {\bibinfo {author} {\bibfnamefont {R.}~\bibnamefont
  {Barbieri}}, \bibinfo {author} {\bibfnamefont {D.}~\bibnamefont {Greco}},
  \bibinfo {author} {\bibfnamefont {R.}~\bibnamefont {Rattazzi}},\ and\
  \bibinfo {author} {\bibfnamefont {A.}~\bibnamefont {Wulzer}},\ }\bibfield
  {title} {\bibinfo {title} {{The Composite Twin Higgs scenario}},\ }\href
  {https://doi.org/10.1007/JHEP08(2015)161} {\bibfield  {journal} {\bibinfo
  {journal} {JHEP}\ }\textbf {\bibinfo {volume} {08}},\ \bibinfo {pages}
  {161}},\ \Eprint {https://arxiv.org/abs/1501.07803} {arXiv:1501.07803
  [hep-ph]} \BibitemShut {NoStop}%
\bibitem [{\citenamefont {Chacko}\ \emph {et~al.}(2017)\citenamefont {Chacko},
  \citenamefont {Craig}, \citenamefont {Fox},\ and\ \citenamefont
  {Harnik}}]{Chacko:2016hvu}%
  \BibitemOpen
  \bibfield  {author} {\bibinfo {author} {\bibfnamefont {Z.}~\bibnamefont
  {Chacko}}, \bibinfo {author} {\bibfnamefont {N.}~\bibnamefont {Craig}},
  \bibinfo {author} {\bibfnamefont {P.~J.}\ \bibnamefont {Fox}},\ and\ \bibinfo
  {author} {\bibfnamefont {R.}~\bibnamefont {Harnik}},\ }\bibfield  {title}
  {\bibinfo {title} {{Cosmology in Mirror Twin Higgs and Neutrino Masses}},\
  }\href {https://doi.org/10.1007/JHEP07(2017)023} {\bibfield  {journal}
  {\bibinfo  {journal} {JHEP}\ }\textbf {\bibinfo {volume} {07}},\ \bibinfo
  {pages} {023}},\ \Eprint {https://arxiv.org/abs/1611.07975} {arXiv:1611.07975
  [hep-ph]} \BibitemShut {NoStop}%
\bibitem [{\citenamefont {Chacko}\ \emph {et~al.}(2018)\citenamefont {Chacko},
  \citenamefont {Curtin}, \citenamefont {Geller},\ and\ \citenamefont
  {Tsai}}]{Chacko:2018vss}%
  \BibitemOpen
  \bibfield  {author} {\bibinfo {author} {\bibfnamefont {Z.}~\bibnamefont
  {Chacko}}, \bibinfo {author} {\bibfnamefont {D.}~\bibnamefont {Curtin}},
  \bibinfo {author} {\bibfnamefont {M.}~\bibnamefont {Geller}},\ and\ \bibinfo
  {author} {\bibfnamefont {Y.}~\bibnamefont {Tsai}},\ }\bibfield  {title}
  {\bibinfo {title} {{Cosmological Signatures of a Mirror Twin Higgs}},\ }\href
  {https://doi.org/10.1007/JHEP09(2018)163} {\bibfield  {journal} {\bibinfo
  {journal} {JHEP}\ }\textbf {\bibinfo {volume} {09}},\ \bibinfo {pages}
  {163}},\ \Eprint {https://arxiv.org/abs/1803.03263} {arXiv:1803.03263
  [hep-ph]} \BibitemShut {NoStop}%
\bibitem [{\citenamefont {Batell}\ \emph {et~al.}(2026)\citenamefont {Batell},
  \citenamefont {Low}, \citenamefont {Neil},\ and\ \citenamefont
  {Verhaaren}}]{Batell:2022tif}%
  \BibitemOpen
  \bibfield  {author} {\bibinfo {author} {\bibfnamefont {B.}~\bibnamefont
  {Batell}}, \bibinfo {author} {\bibfnamefont {M.}~\bibnamefont {Low}},
  \bibinfo {author} {\bibfnamefont {E.~T.}\ \bibnamefont {Neil}},\ and\
  \bibinfo {author} {\bibfnamefont {C.~B.}\ \bibnamefont {Verhaaren}},\
  }\bibfield  {title} {\bibinfo {title} {{Review of neutral naturalness}},\
  }\href {https://doi.org/10.1016/j.physrep.2025.12.001} {\bibfield  {journal}
  {\bibinfo  {journal} {Phys. Rept.}\ }\textbf {\bibinfo {volume} {1165}},\
  \bibinfo {pages} {1} (\bibinfo {year} {2026})},\ \Eprint
  {https://arxiv.org/abs/2203.05531} {arXiv:2203.05531 [hep-ph]} \BibitemShut
  {NoStop}%
\bibitem [{\citenamefont {Juknevich}(2010)}]{Juknevich:2009gg}%
  \BibitemOpen
  \bibfield  {author} {\bibinfo {author} {\bibfnamefont {J.~E.}\ \bibnamefont
  {Juknevich}},\ }\bibfield  {title} {\bibinfo {title} {{Pure-glue hidden
  valleys through the Higgs portal}},\ }\href
  {https://doi.org/10.1007/JHEP08(2010)121} {\bibfield  {journal} {\bibinfo
  {journal} {JHEP}\ }\textbf {\bibinfo {volume} {08}},\ \bibinfo {pages}
  {121}},\ \Eprint {https://arxiv.org/abs/0911.5616} {arXiv:0911.5616 [hep-ph]}
  \BibitemShut {NoStop}%
\bibitem [{\citenamefont {Batz}\ \emph {et~al.}(2024)\citenamefont {Batz},
  \citenamefont {Cohen}, \citenamefont {Curtin}, \citenamefont {Gemmell},\ and\
  \citenamefont {Kribs}}]{Batz:2023zef}%
  \BibitemOpen
  \bibfield  {author} {\bibinfo {author} {\bibfnamefont {A.}~\bibnamefont
  {Batz}}, \bibinfo {author} {\bibfnamefont {T.}~\bibnamefont {Cohen}},
  \bibinfo {author} {\bibfnamefont {D.}~\bibnamefont {Curtin}}, \bibinfo
  {author} {\bibfnamefont {C.}~\bibnamefont {Gemmell}},\ and\ \bibinfo {author}
  {\bibfnamefont {G.~D.}\ \bibnamefont {Kribs}},\ }\bibfield  {title} {\bibinfo
  {title} {{Dark sector glueballs at the LHC}},\ }\href
  {https://doi.org/10.1007/JHEP04(2024)070} {\bibfield  {journal} {\bibinfo
  {journal} {JHEP}\ }\textbf {\bibinfo {volume} {04}},\ \bibinfo {pages}
  {070}},\ \Eprint {https://arxiv.org/abs/2310.13731} {arXiv:2310.13731
  [hep-ph]} \BibitemShut {NoStop}%
\bibitem [{\citenamefont {Batell}\ \emph {et~al.}(2023)\citenamefont {Batell},
  \citenamefont {Ghosh}, \citenamefont {Han},\ and\ \citenamefont
  {Xie}}]{Batell:2022ogj}%
  \BibitemOpen
  \bibfield  {author} {\bibinfo {author} {\bibfnamefont {B.}~\bibnamefont
  {Batell}}, \bibinfo {author} {\bibfnamefont {T.}~\bibnamefont {Ghosh}},
  \bibinfo {author} {\bibfnamefont {T.}~\bibnamefont {Han}},\ and\ \bibinfo
  {author} {\bibfnamefont {K.}~\bibnamefont {Xie}},\ }\bibfield  {title}
  {\bibinfo {title} {{Heavy neutral leptons at the Electron-Ion Collider}},\
  }\href {https://doi.org/10.1007/JHEP03(2023)020} {\bibfield  {journal}
  {\bibinfo  {journal} {JHEP}\ }\textbf {\bibinfo {volume} {03}},\ \bibinfo
  {pages} {020}},\ \Eprint {https://arxiv.org/abs/2210.09287} {arXiv:2210.09287
  [hep-ph]} \BibitemShut {NoStop}%
\bibitem [{\citenamefont {Davoudiasl}\ \emph {et~al.}(2023)\citenamefont
  {Davoudiasl}, \citenamefont {Marcarelli},\ and\ \citenamefont
  {Neil}}]{Davoudiasl:2023pkq}%
  \BibitemOpen
  \bibfield  {author} {\bibinfo {author} {\bibfnamefont {H.}~\bibnamefont
  {Davoudiasl}}, \bibinfo {author} {\bibfnamefont {R.}~\bibnamefont
  {Marcarelli}},\ and\ \bibinfo {author} {\bibfnamefont {E.~T.}\ \bibnamefont
  {Neil}},\ }\bibfield  {title} {\bibinfo {title} {{Displaced signals of hidden
  vectors at the Electron-Ion Collider}},\ }\href
  {https://doi.org/10.1103/PhysRevD.108.075017} {\bibfield  {journal} {\bibinfo
   {journal} {Phys. Rev. D}\ }\textbf {\bibinfo {volume} {108}},\ \bibinfo
  {pages} {075017} (\bibinfo {year} {2023})},\ \Eprint
  {https://arxiv.org/abs/2307.00102} {arXiv:2307.00102 [hep-ph]} \BibitemShut
  {NoStop}%
\bibitem [{\citenamefont {Balkin}\ \emph {et~al.}(2024)\citenamefont {Balkin},
  \citenamefont {Hen}, \citenamefont {Li}, \citenamefont {Liu}, \citenamefont
  {Ma}, \citenamefont {Soreq},\ and\ \citenamefont
  {Williams}}]{Balkin:2023gya}%
  \BibitemOpen
  \bibfield  {author} {\bibinfo {author} {\bibfnamefont {R.}~\bibnamefont
  {Balkin}}, \bibinfo {author} {\bibfnamefont {O.}~\bibnamefont {Hen}},
  \bibinfo {author} {\bibfnamefont {W.}~\bibnamefont {Li}}, \bibinfo {author}
  {\bibfnamefont {H.}~\bibnamefont {Liu}}, \bibinfo {author} {\bibfnamefont
  {T.}~\bibnamefont {Ma}}, \bibinfo {author} {\bibfnamefont {Y.}~\bibnamefont
  {Soreq}},\ and\ \bibinfo {author} {\bibfnamefont {M.}~\bibnamefont
  {Williams}},\ }\bibfield  {title} {\bibinfo {title} {{Probing axion-like
  particles at the Electron-Ion Collider}},\ }\href
  {https://doi.org/10.1007/JHEP02(2024)123} {\bibfield  {journal} {\bibinfo
  {journal} {JHEP}\ }\textbf {\bibinfo {volume} {02}},\ \bibinfo {pages}
  {123}},\ \Eprint {https://arxiv.org/abs/2310.08827} {arXiv:2310.08827
  [hep-ph]} \BibitemShut {NoStop}%
\bibitem [{\citenamefont {Davoudiasl}\ and\ \citenamefont
  {Liu}(2025)}]{Davoudiasl:2025rpn}%
  \BibitemOpen
  \bibfield  {author} {\bibinfo {author} {\bibfnamefont {H.}~\bibnamefont
  {Davoudiasl}}\ and\ \bibinfo {author} {\bibfnamefont {H.}~\bibnamefont
  {Liu}},\ }\bibfield  {title} {\bibinfo {title} {{Electron-ion collider as a
  discovery tool for invisible dark bosons}},\ }\href
  {https://doi.org/10.1103/gtsf-24x4} {\bibfield  {journal} {\bibinfo
  {journal} {Phys. Rev. D}\ }\textbf {\bibinfo {volume} {112}},\ \bibinfo
  {pages} {075001} (\bibinfo {year} {2025})},\ \Eprint
  {https://arxiv.org/abs/2505.08871} {arXiv:2505.08871 [hep-ph]} \BibitemShut
  {NoStop}%
\bibitem [{\citenamefont {Balkin}\ \emph {et~al.}(2025)\citenamefont {Balkin},
  \citenamefont {Coren}, \citenamefont {Jentsch}, \citenamefont {Liu},
  \citenamefont {Ovchynnikov}, \citenamefont {Soreq},\ and\ \citenamefont
  {Trifinopoulos}}]{Balkin:2025rtc}%
  \BibitemOpen
  \bibfield  {author} {\bibinfo {author} {\bibfnamefont {R.}~\bibnamefont
  {Balkin}}, \bibinfo {author} {\bibfnamefont {T.}~\bibnamefont {Coren}},
  \bibinfo {author} {\bibfnamefont {A.}~\bibnamefont {Jentsch}}, \bibinfo
  {author} {\bibfnamefont {H.}~\bibnamefont {Liu}}, \bibinfo {author}
  {\bibfnamefont {M.}~\bibnamefont {Ovchynnikov}}, \bibinfo {author}
  {\bibfnamefont {Y.}~\bibnamefont {Soreq}},\ and\ \bibinfo {author}
  {\bibfnamefont {S.}~\bibnamefont {Trifinopoulos}},\ }\bibfield  {title}
  {\bibinfo {title} {{Braking protons at the EIC: from invisible meson decay to
  new physics searches}},\ }\href@noop {} {\  (\bibinfo {year} {2025})},\
  \Eprint {https://arxiv.org/abs/2601.00068} {arXiv:2601.00068 [hep-ph]}
  \BibitemShut {NoStop}%
\bibitem [{\citenamefont {Balkin}\ \emph {et~al.}(2026)\citenamefont {Balkin},
  \citenamefont {Coren}, \citenamefont {Jentsch}, \citenamefont {Liu},
  \citenamefont {Ovchynnikov}, \citenamefont {Soreq},\ and\ \citenamefont
  {Trifinopoulos}}]{Balkin:2026whv}%
  \BibitemOpen
  \bibfield  {author} {\bibinfo {author} {\bibfnamefont {R.}~\bibnamefont
  {Balkin}}, \bibinfo {author} {\bibfnamefont {T.}~\bibnamefont {Coren}},
  \bibinfo {author} {\bibfnamefont {A.}~\bibnamefont {Jentsch}}, \bibinfo
  {author} {\bibfnamefont {H.}~\bibnamefont {Liu}}, \bibinfo {author}
  {\bibfnamefont {M.}~\bibnamefont {Ovchynnikov}}, \bibinfo {author}
  {\bibfnamefont {Y.}~\bibnamefont {Soreq}},\ and\ \bibinfo {author}
  {\bibfnamefont {S.}~\bibnamefont {Trifinopoulos}},\ }\bibfield  {title}
  {\bibinfo {title} {{On Exclusive Coherent Production of Bosons in
  Electron-Proton Collisions}},\ }\href@noop {} {\  (\bibinfo {year} {2026})},\
  \Eprint {https://arxiv.org/abs/2604.08667} {arXiv:2604.08667 [hep-ph]}
  \BibitemShut {NoStop}%
\bibitem [{\citenamefont {DeGrand}\ and\ \citenamefont
  {Neil}(2020)}]{DeGrand:2019vbx}%
  \BibitemOpen
  \bibfield  {author} {\bibinfo {author} {\bibfnamefont {T.}~\bibnamefont
  {DeGrand}}\ and\ \bibinfo {author} {\bibfnamefont {E.~T.}\ \bibnamefont
  {Neil}},\ }\bibfield  {title} {\bibinfo {title} {{Repurposing lattice QCD
  results for composite phenomenology}},\ }\href
  {https://doi.org/10.1103/PhysRevD.101.034504} {\bibfield  {journal} {\bibinfo
   {journal} {Phys. Rev. D}\ }\textbf {\bibinfo {volume} {101}},\ \bibinfo
  {pages} {034504} (\bibinfo {year} {2020})},\ \bibinfo {note} {[Erratum:
  Phys.Rev.D 112, 079902 (2025)]},\ \Eprint {https://arxiv.org/abs/1910.08561}
  {arXiv:1910.08561 [hep-ph]} \BibitemShut {NoStop}%
\bibitem [{\citenamefont {McKeen}\ \emph {et~al.}(2018)\citenamefont {McKeen},
  \citenamefont {Nelson}, \citenamefont {Reddy},\ and\ \citenamefont
  {Zhou}}]{McKeen:2018xwc}%
  \BibitemOpen
  \bibfield  {author} {\bibinfo {author} {\bibfnamefont {D.}~\bibnamefont
  {McKeen}}, \bibinfo {author} {\bibfnamefont {A.~E.}\ \bibnamefont {Nelson}},
  \bibinfo {author} {\bibfnamefont {S.}~\bibnamefont {Reddy}},\ and\ \bibinfo
  {author} {\bibfnamefont {D.}~\bibnamefont {Zhou}},\ }\bibfield  {title}
  {\bibinfo {title} {{Neutron stars exclude light dark baryons}},\ }\href
  {https://doi.org/10.1103/PhysRevLett.121.061802} {\bibfield  {journal}
  {\bibinfo  {journal} {Phys. Rev. Lett.}\ }\textbf {\bibinfo {volume} {121}},\
  \bibinfo {pages} {061802} (\bibinfo {year} {2018})},\ \Eprint
  {https://arxiv.org/abs/1802.08244} {arXiv:1802.08244 [hep-ph]} \BibitemShut
  {NoStop}%
\bibitem [{\citenamefont {Carloni}\ and\ \citenamefont
  {Sjostrand}(2010)}]{Carloni:2010tw}%
  \BibitemOpen
  \bibfield  {author} {\bibinfo {author} {\bibfnamefont {L.}~\bibnamefont
  {Carloni}}\ and\ \bibinfo {author} {\bibfnamefont {T.}~\bibnamefont
  {Sjostrand}},\ }\bibfield  {title} {\bibinfo {title} {{Visible Effects of
  Invisible Hidden Valley Radiation}},\ }\href
  {https://doi.org/10.1007/JHEP09(2010)105} {\bibfield  {journal} {\bibinfo
  {journal} {JHEP}\ }\textbf {\bibinfo {volume} {09}},\ \bibinfo {pages}
  {105}},\ \Eprint {https://arxiv.org/abs/1006.2911} {arXiv:1006.2911 [hep-ph]}
  \BibitemShut {NoStop}%
\bibitem [{\citenamefont {Carloni}\ \emph {et~al.}(2011)\citenamefont
  {Carloni}, \citenamefont {Rathsman},\ and\ \citenamefont
  {Sjostrand}}]{Carloni:2011kk}%
  \BibitemOpen
  \bibfield  {author} {\bibinfo {author} {\bibfnamefont {L.}~\bibnamefont
  {Carloni}}, \bibinfo {author} {\bibfnamefont {J.}~\bibnamefont {Rathsman}},\
  and\ \bibinfo {author} {\bibfnamefont {T.}~\bibnamefont {Sjostrand}},\
  }\bibfield  {title} {\bibinfo {title} {{Discerning Secluded Sector gauge
  structures}},\ }\href {https://doi.org/10.1007/JHEP04(2011)091} {\bibfield
  {journal} {\bibinfo  {journal} {JHEP}\ }\textbf {\bibinfo {volume} {04}},\
  \bibinfo {pages} {091}},\ \Eprint {https://arxiv.org/abs/1102.3795}
  {arXiv:1102.3795 [hep-ph]} \BibitemShut {NoStop}%
\bibitem [{\citenamefont {Bierlich}\ \emph {et~al.}(2022)\citenamefont
  {Bierlich} \emph {et~al.}}]{Bierlich:2022pfr}%
  \BibitemOpen
  \bibfield  {author} {\bibinfo {author} {\bibfnamefont {C.}~\bibnamefont
  {Bierlich}} \emph {et~al.},\ }\bibfield  {title} {\bibinfo {title} {{A
  comprehensive guide to the physics and usage of PYTHIA 8.3}},\ }\href
  {https://doi.org/10.21468/SciPostPhysCodeb.8} {\bibfield  {journal} {\bibinfo
   {journal} {SciPost Phys. Codeb.}\ }\textbf {\bibinfo {volume} {2022}},\
  \bibinfo {pages} {8} (\bibinfo {year} {2022})},\ \Eprint
  {https://arxiv.org/abs/2203.11601} {arXiv:2203.11601 [hep-ph]} \BibitemShut
  {NoStop}%
\bibitem [{\citenamefont {Helm}(1956)}]{Helm:1956zz}%
  \BibitemOpen
  \bibfield  {author} {\bibinfo {author} {\bibfnamefont {R.~H.}\ \bibnamefont
  {Helm}},\ }\bibfield  {title} {\bibinfo {title} {{Inelastic and Elastic
  Scattering of 187-Mev Electrons from Selected Even-Even Nuclei}},\ }\href
  {https://doi.org/10.1103/PhysRev.104.1466} {\bibfield  {journal} {\bibinfo
  {journal} {Phys. Rev.}\ }\textbf {\bibinfo {volume} {104}},\ \bibinfo {pages}
  {1466} (\bibinfo {year} {1956})}\BibitemShut {NoStop}%
\bibitem [{\citenamefont {Aubert}\ \emph {et~al.}(2002)\citenamefont {Aubert}
  \emph {et~al.}}]{BaBar:2001yhh}%
  \BibitemOpen
  \bibfield  {author} {\bibinfo {author} {\bibfnamefont {B.}~\bibnamefont
  {Aubert}} \emph {et~al.} (\bibinfo {collaboration} {BaBar}),\ }\bibfield
  {title} {\bibinfo {title} {{The BaBar detector}},\ }\href
  {https://doi.org/10.1016/S0168-9002(01)02012-5} {\bibfield  {journal}
  {\bibinfo  {journal} {Nucl. Instrum. Meth. A}\ }\textbf {\bibinfo {volume}
  {479}},\ \bibinfo {pages} {1} (\bibinfo {year} {2002})},\ \Eprint
  {https://arxiv.org/abs/hep-ex/0105044} {arXiv:hep-ex/0105044} \BibitemShut
  {NoStop}%
\bibitem [{\citenamefont {Lees}\ \emph {et~al.}(2015)\citenamefont {Lees} \emph
  {et~al.}}]{BaBar:2015jvu}%
  \BibitemOpen
  \bibfield  {author} {\bibinfo {author} {\bibfnamefont {J.~P.}\ \bibnamefont
  {Lees}} \emph {et~al.} (\bibinfo {collaboration} {BaBar}),\ }\bibfield
  {title} {\bibinfo {title} {{Search for Long-Lived Particles in $e^+e^-$
  Collisions}},\ }\href {https://doi.org/10.1103/PhysRevLett.114.171801}
  {\bibfield  {journal} {\bibinfo  {journal} {Phys. Rev. Lett.}\ }\textbf
  {\bibinfo {volume} {114}},\ \bibinfo {pages} {171801} (\bibinfo {year}
  {2015})},\ \Eprint {https://arxiv.org/abs/1502.02580} {arXiv:1502.02580
  [hep-ex]} \BibitemShut {NoStop}%
\bibitem [{\citenamefont {Adachi}\ \emph {et~al.}(2023)\citenamefont {Adachi}
  \emph {et~al.}}]{Belle-II:2023ueh}%
  \BibitemOpen
  \bibfield  {author} {\bibinfo {author} {\bibfnamefont {I.}~\bibnamefont
  {Adachi}} \emph {et~al.} (\bibinfo {collaboration} {Belle-II}),\ }\bibfield
  {title} {\bibinfo {title} {{Search for a long-lived spin-0 mediator in
  b{\textrightarrow}s transitions at the Belle II experiment}},\ }\href
  {https://doi.org/10.1103/PhysRevD.108.L111104} {\bibfield  {journal}
  {\bibinfo  {journal} {Phys. Rev. D}\ }\textbf {\bibinfo {volume} {108}},\
  \bibinfo {pages} {L111104} (\bibinfo {year} {2023})},\ \Eprint
  {https://arxiv.org/abs/2306.02830} {arXiv:2306.02830 [hep-ex]} \BibitemShut
  {NoStop}%
\bibitem [{\citenamefont {Jaeckel}\ and\ \citenamefont
  {Phan}(2024)}]{Jaeckel:2023huy}%
  \BibitemOpen
  \bibfield  {author} {\bibinfo {author} {\bibfnamefont {J.}~\bibnamefont
  {Jaeckel}}\ and\ \bibinfo {author} {\bibfnamefont {A.~V.}\ \bibnamefont
  {Phan}},\ }\bibfield  {title} {\bibinfo {title} {{Searching dark photons
  using displaced vertices at Belle II {\textemdash} with backgrounds}},\
  }\href {https://doi.org/10.1007/JHEP08(2024)062} {\bibfield  {journal}
  {\bibinfo  {journal} {JHEP}\ }\textbf {\bibinfo {volume} {08}},\ \bibinfo
  {pages} {062}},\ \Eprint {https://arxiv.org/abs/2312.12522} {arXiv:2312.12522
  [hep-ph]} \BibitemShut {NoStop}%
\end{thebibliography}%

\end{document}